\documentclass[pra,prl,floatfix,twocolumn, showpacs]{revtex4}%
\usepackage{mathrsfs}
\usepackage{graphicx}
\usepackage{slashbox}
\usepackage{amssymb}
\usepackage{verbatim}
\usepackage{amsmath}
\usepackage{amsfonts}%
\providecommand{\U}[1]{\protect\rule{.1in}{.1in}}
\begin{document}
\title{Correlation in states of two identical particles}
\author{D.L. Zhou}
\affiliation{Beijing National Laboratory for Condensed Matter Physics}
\affiliation{Institute of Physics, Chinese Academy of Sciences, Beijing 100080, China}

\begin{abstract}
We identify the correlation in a state of two identical particles as
the residual information beyond what is already contained in the
$1$-particle reduced density matrix, and propose a correlation
measure based on the maximum entropy principle. We obtain the
analytical results of the correlation measure, which make it
computable for arbitrary two-particle states. We also show that the
degrees of correlation in the same two-particle states with
different particle types will decrease in the following order:
bosons, fermions, and distinguishable particles.
\end{abstract}

\maketitle

\textit{Introduction.} --- The classification and characterization
of correlations in a multi-party quantum state is a fundamental
problem in quantum many particle physics and quantum information
science \cite{Linden,Zhou1}. Although the theory on correlations in
a state of more than two distinguishable particles is still under
developing, the correlation in a state of two distinguishable
particles is believed to be completely understood. A state of two
distinguishable particles is uncorrelated if and only if it is a
product state. The degree of correlation between the first particle
and the second particle in a state of two distinguishable particles
is shown to be the mutual entropy of the state
\cite{Linden,Groisman,Schumacher}. However, these results can not be
generalized directly to the case of a state for two identical
particles. For example, we have no idea of what is a product state
for two identical particles.

The concept of identical particles in quantum mechanics is
essentially different from that in classical mechanics. Therefore
the theory on correlations in a state of identical particles must
consider the feature of identical particles. Particularly, identical
particles in quantum mechanics are absolutely indistinguishable.
Thus, for a system composed by two identical particles, the term
like ``the correlation between the first particle and the second
particle'' loses its meaning. Naturally we ask the following
fundamental question: \textit{what is the correlation in a quantum
state of two identical particles?}

Based on the consideration of practical use of the state of
identical particles as an entanglement resources, a useful approach
\cite{ZW02,GF02,WV03,S03} to solve this problem is proposed, whose
basic idea is as follows. First, two distinguishable subsets of
orthogonal modes are specified (here ``mode'' is abbreviated for
``single particle state''). Then the correlation between the first
subset of modes and the second subset of modes can be defined
normally. Obviously, this type of correlation strongly depends on
the specified subsets of modes, and therefore it is not an intrinsic
property of the state.

Another instructive approach \cite{SCKML01,PY01,YZLL01,GM04} is
built on the mathematical similarity of the structure of pure states
between two distinguishable particles and two identical particles.
As is well known, the Schmidt decomposition of a pure state of two
distinguishable particles plays an important role in the
characterization of its entanglement. Fortunately, the generalized
Schmidt decomposition for pure states of two identical particles are
discovered. The von Neaumann entropy of the normalized $1$-particle
reduced density matrix is suggested as an entanglement measure for a
pure state of two identical particles \cite{PY01}.

Our correlation measure for a state of two identical particles is
built on the maximum entropy principle: the least biased state on
the basis of partial knowledge is the state with the maximum entropy
under the constraint by the partial knowledge. This principle was
first proposed to be used as the foundation of statistical mechanics
by Jaynes in 1950s \cite{Jaynes}. Remarkably, it also found
important applications in characterizing different types of
correlations in probability distribution of random variables
\cite{Schneidman,Amari,Jost} or quantum states of distinguishable
particles \cite{Linden,Zhou2}.

In this Letter, we propose a correlation measure for a quantum state
of two identical particles based on the maximum entropy principle.
Here the correlation in a state of two identical particles is
identified as \textit{the residual information beyond what is
already contained in the $1$-particle reduced density matrix}.

Based on our correlation measure, we show that the degree of
correlation is not only determined by the two-particle quantum
state, but also essentially influenced by the types of the two
particles. Actually, the degrees of correlation in the same
two-particle states with different particle types will decrease in
the following order: bosons, fermions, and distinguishable
particles.


\textit{Definition.} --- The quantum state of a system composed by
two identical particles is specified by a two-particle density
matrix $\sigma^{(2)}$, whose $1$-particle reduced density matrix
$\sigma^{(1)}$ \cite{Yang} is defined by
\begin{equation}
\sigma^{(1)}_{\eta \tau}\equiv \mathrm{Tr}\big(
a_{\eta}\sigma^{(1)}a_{\tau}^{\dagger}\big) =\mathrm{Tr}\big(
a_{\eta}\sigma^{(2)}a_{\tau}^{\dagger}\big)  ,
\label{eq1}%
\end{equation}
where the operator $a_{\eta}$ and $a_{\tau}^{\dagger}$ are the
annihilation operator of mode $\eta$ and the creation operator of
mode $\tau$ respectively. The $1$-particle reduced density matrix
$\sigma^{(1)}$ tells us the average particle number in any mode for
the state $\sigma^{(2)}$, and can be obtained by measuring
$1$-particle observables.

Now we apply the maximum entropy principle to define a correlation
measure for the state $\sigma^{(2)}$ as follows. First we define a
set of two-particle states with the same $1$-particle reduced
density matrix as
that of the state $\sigma^{( 2)}$, \textit{i.e.},%

\begin{equation}
D_{1}\big(\sigma^{(2)}\big)=\big\{\rho^{(2)} \;\big|\; \rho^{(1)
}=\sigma^{(1)}\big\}. \label{eq2}%
\end{equation}
Next we find out, among the states in the set
$D_{1}\big(\sigma^{(2)}\big)$, the state $\sigma_{1}^{(2)}$ that
takes the maximal entropy, \textit{i.e.},
\begin{equation}
\sigma_{1}^{(2)}  ={\arg\max}_{\rho^{\left(  2\right)  }\in
D_{1}(\sigma^{(2)})}S\big(\rho^{(2)}\big), \label{eq3}%
\end{equation}
where the von Neaumann entropy $S\big( \rho\big)
=-\mathrm{Tr}\rho\ln
\rho$. Then a correlation measure for the state $\sigma^{(2)}$ is defined as%
\begin{equation}
C_{2}\big(\sigma^{(2)}\big)  =S\big(  \sigma_{1}^{(2)}\big) -S\big(
\sigma^{(2)}\big)  . \label{eq4}%
\end{equation}

There are two essential elements in defining the above measure. One
is to associate the characterization of non-correlation property of
the state $\sigma^{(2)}$ with the $1$-particle reduced density
matrix $\sigma^{(1)}$. The other is to realize that a state with
more correlation has less entropy. Thus the correlation measure of
the state $\sigma^{(2)}$ is equal to the entropy difference between
the state $\sigma^{(2)}_{1}$ and the state $\sigma^{(2)}$, where the
state $\sigma^{(2)}_{1}$ is the state with maximal entropy and with
the same $1$-particle reduced density matrix as $\sigma^{(1)}$.

Obviously, the correlation measure $C_{2}(\sigma^{(2)}_{1})=0$,
namely, the state $\sigma^{(2)}_{1}$ is a uncorrelated state of two
identical particles. Thus, the exact role played by the maximum
entropy principle is to provide us a mathematical tool to define the
uncorrelated state of two identical particles by Eq. (\ref{eq3}).

\textit{Analytic results.} --- Since the state $\sigma^{(2)}_{1}$ is
determined by a constrained optimization defined by Eqs. (\ref{eq2})
and (\ref{eq3}), we obtain an analytic result of the state
$\sigma^{(2)}_1$ by applying the standard Lagrange multipliers
method. A simple calculation will show that

\begin{equation}
\sigma_{1}^{(2)}=\exp\big(
\sum_{\mu}\gamma_{\mu}\hat{n}_{\mu}\big)  =\prod_{\mu}x_{\mu}^{\hat{n}_{\mu}}, \label{eq5}%
\end{equation}
where the parameters $\gamma_{\mu}$ are the Lagrange multipliers in
the diagonal representation, the parameters $x_{\mu}=\exp(
\gamma_{\mu}) $, and the total particle number $\hat{n}=\sum_{\mu}%
\hat{n}_{\mu}=2$. The unknown parameters $\gamma_{\mu}$ and
$x_{\mu}$ are determined by the following equation
\begin{equation}
\sigma^{(1)}_{1}=\sigma^{(1)}.\label{eq6}
\end{equation}

Eqs. (\ref{eq5}) and (\ref{eq6}) imply that the diagonal modes $\mu$
of the state $\sigma^{(2)}_{1}$ are the same as the diagonal modes
of the state $\sigma^{(1)}$. Thus, in practice, we can first get the
diagonal representation of the state $\sigma^{(1)}$ and determine
the modes $\mu$. Then we only need to solve Eq. (\ref{eq6}) in the
diagonal representation.

All the above discussions are valid for both bosons and fermions. As
is well known, there is a basic distinction between bosons and
fermions. For bosons, the occupation number of a single mode can be
arbitrary; while, for fermions, the occupation number of a single
mode is $0$ or $1$. The distinction, together with the total
particle number requirement $\hat{n}=2$, makes Eq. (\ref{eq6})
become
\begin{eqnarray}
x_{\mu}\sum_{\nu}x_{\nu}\pm x_{\mu}^{2}=\sigma^{(1)}_{\mu\mu},
\label{eq7}
\end{eqnarray}
where the sign $+$ is for bosons, and the sign $-$ is for fermions.

Two direct results can be derived from the above analytical results.
First, a state of two identical particles is uncorrelated if and
only if it can be written in the form given by Eq. (\ref{eq5}).
Second, Eq. (\ref{eq7}) makes the correlation measure defined by Eq.
(\ref{eq4}) become computable for arbitrary states of two identical
particles.

\textit{Indistinguishable and distinguishable.} --- For two
particles of different types, \textit{e.g.}, with different masses
or different charges, the two particles are absolutely
distinguishable. However, for two identical particles, if the modes
can be divided into two subsets $\{A\alpha\}$ and $\{B\beta\}$
respectively, and $\hat{n}_{A}=\hat{n}_{B}=1$, the two identical
particles become effectively distinguishable, \textit{i.e.}, we can
call one particle as particle $A$, the other as particle $B$.

Under the condition $\hat{n}_{A}=\hat{n}_{B}=1$, it is easy to prove
that the $1$-particle reduced density matrix $\sigma^{(1)}$ is a
two-block diagonal matrix, with one block corresponding to particle
$A$ and the other block corresponding to particle $B$, denoted by
$\sigma^{(A)}$ and $\sigma^{(B)}$ respectively. Let us further
denote the modes for the diagonal representation of $\sigma^{(A)}$
and $\sigma^{(B)}$ are $\{A\mu\}$ and $\{B\nu\}$. Now Eq.
(\ref{eq6}) becomes $ x_{A\mu}\sum_{\nu}x_{B\nu}  =
\sigma^{(A)}_{\mu\mu}$ , and $x_{B\nu}\sum_{\mu}x_{A\mu}
=\sigma^{(B)}_{\nu\nu}$.

The normalization condition of the state $\sigma^{(2)}_{1}$ requires
that $\sum_{\mu\nu}x_{A\mu} x_{B\nu}=1$. Through a few steps of
calculation, as expected, we can prove that the correlation measure
is equal to the mutual entropy, \textit{i.e.},
\begin{equation}
C_2(\sigma^{(2)})=S(\sigma^{(A)})+S(\sigma^{(B)})-S(\sigma^{(2)}).
\label{eq8}
\end{equation}

In the above discussion, there is no differences for bosons and
fermions, which is due to the condition $\hat{n}_{A}=\hat{n}_{B}=1$.

\textit{Pure states and Schmidt decomposition.} --- Correlation in
pure states of two identical particles are sometimes called quantum
correlation. When two identical particles becomes effectively
distinguishable, it is also called quantum entanglement. For a
two-particle pure state $\sigma^{(2)}$, the entropy $S\big(
\sigma^{(2)}\big) =0$.

Further more, pure states of two identical particles can always be
written in the Schmidt decomposition forms, whether the two
identical particles are bosons, fermions, or effectively
distinguishable. These forms can be explicitly written as
$\sum_{\mu} \sqrt { {\sigma^{(1)}_{\mu\mu}}/ {2} } |2_{\mu}\rangle$
for bosons, $\sum_{\mu} \sqrt { {\sigma^{(1)}_{{2\mu}\,{2\mu}}} }
|1_{2\mu-1} 1_{2\mu}\rangle$ for fermions, and  $\sum_{\mu} \sqrt {
{\sigma^{(A)}_{\mu\mu}} } |1_{A\mu} 1_{B\mu}\rangle$ for effectively
distinguishable particles. In the above forms,
$\sigma^{(1)}_{{2\mu}\,{2\mu}}=\sigma^{(1)}_{{2\mu-1}\,{2\mu-1}}$
for fermions, and $\sigma^{(A)}_{\mu\mu}=\sigma^{(B)}_{\mu\mu}$ for
effectively distinguishable particles. The advantages of these forms
are that they give directly all the information of the $1$-particle
reduced density matrix $\sigma^{(1)}$. Thus, for pure states of two
identical particles, the correlation measure defined by Eq.
(\ref{eq4}) can be obtained by directly applying Eqs. (\ref{eq5})
and (\ref{eq7}).

It is easy to show that a pure two-particle state is uncorrelated if
and only if the Schmidt number is $1$. In other words, for pure
states, two distinguishable particles or two fermions are
uncorrelated if and only if they occupy two orthogonal modes
respectively, while two bosons are uncorrelated if and only if they
occupy the same mode.

\textit{Correlation inequality.} --- In Eq. (\ref{eq4}), the degree
of correlation in a two-particle state is defined by the uncertainty
decrease induced by the correlation. This definition implies that
the degrees of different types of correlations are related with the
same quantity, uncertainty. Thus, it is possible and reasonable to
compare the degrees of different types of correlations. For example,
in Ref. \cite{Zhou2} we have shown that the degree of  the total
correlation in an $n$-partite quantum state is the sum of the
degrees of all the irreducible $k$-party ($2\le k\le n$)
correlations. Now we consider the comparison of the degrees of
correlations in the states for systems composed by different types
of particles.

Notice that, for any state $\sigma_{D}%
^{(2)}$ of two  effectively distinguishable particles, there exist
the counterpart states for two fermions and two bosons, which are
obtained by regarding the particle indexes A and B as the indexes of
the different subsets of modes. Here ``the different subsets of
modes'' means that each mode in one subset is orthogonal to all
modes in the other subset.

Let us denote these states with the same form as $\sigma_{D}^{(2)}$
for two effectively distinguishable particles, $\sigma_{F}^{(2)}$
for two fermions, and $\sigma_{B}^{(2)}$ for two bosons,
respectively. It is easy to find that $S( \sigma_{D}^{(2)})=S(
\sigma_{F}^{(2)})=S( \sigma_{B}^{(2)})$ and
$\sigma_{D}^{(1)}=\sigma_{F}^{(1)}=\sigma_{B}^{(1)}$. Let us denote
the basis of diagonal representation of $\sigma^{(1)}$ as $A\mu$ and
$B\nu$. Then the constraints for the two particle Hilbert spaces are
$\hat{n}_{A}=\hat{n}_{B}=1$ for effectively distinguishable
particles, $\hat{n}_{A\mu}, \hat{n}_{B\nu}=0\; \mathrm{or}\; 1$ for
fermions, and $\hat{n}_{A\mu}, \hat{n}_{B\nu}=0 \;\mathrm{or}\; 1
\;\mathrm{or}\; 2$. Comparing these constraints, we know
$D_1(\sigma^{(2)}_{D}) \subseteq D_1(\sigma^{(2)}_{F}) \subseteq
D_1(\sigma^{(2)}_{B})$. Note that, when comparing two states in
different Hilbert spaces, we identify the state in a smaller Hilbert
space with the counterpart state in a larger Hilbert space. Then the
definition (\ref{eq3}) implies that $S( \sigma_{D1}^{(2)})\le S(
\sigma_{F1}^{(2)})\le S( \sigma_{B1}^{(2)})$. Therefore the degrees
of correlations for the three states $\sigma _{D}^{(2)}$,
$\sigma_{F}^{(2)}$, and $\sigma_{B}^{(2)}$ defined above satisfies
the inequality
\begin{equation}
C_{2}\left(  \sigma_{D}^{(2)}\right)  \leq C_{2}\left(  \sigma_{F}%
^{(2)}\right)  \leq C_{2}\left(  \sigma_{B}^{(2)}\right)  . \label{eq22}%
\end{equation}

Eq. (\ref{eq22}) tell us that, in the same quantum state, two
fermions are less correlated than two bosons but more correlated
than two (effectively) distinguishable particles. This result is
originated from different constraints on the Hilbert spaces for
different types of particles, which is explicitly demonstrated by
the following typical examples.

\textit{Examples.} --- In order to demonstrate the power of the
correlation measure, we apply it to analyze the correlations in
three typical two-particle quantum states as listed in Table I. The
values of the correlation measure in Table I are obtained by
applying Eqs. (\ref{eq4}), (\ref{eq5}), and (\ref{eq7}) for bosons
and fermions, and by applying Eq. (\ref{eq8}) for distinguishable
particles respectively.

\begin{table}[h]
\begin{tabular}{|c|c|c|c|}
\hline Typical states & D & F & B\\
\hline
 $|1_{A1} 1_{B1}\rangle$ & $0$& $0$ & $\ln 3$\\
 \hline
 $\frac {1} {\sqrt{2}} (|1_{A1} 1_{B1}\rangle +|1_{A2}
 1_{B2}\rangle) $& $\ln4$ & $\ln6$ &$\ln 10$\\
 \hline
$\frac {1} {2} (|1_{A1} 1_{B1}\rangle \langle 1_{A1} 1_{B1}|
+|1_{A2}
 1_{B2}\rangle \langle 1_{A2} 1_{B2}|)$& $\ln 2$ &$\ln3$ & $\ln5$\\
 \hline
\end{tabular}
\caption{Correlation measures in typical states. Here D, F, B are
abbreviated for distinguishable particles, fermions, and bosons
respectively.}
\end{table}

As expected,  the inequality (\ref{eq22}) is satisfied for the three
states in Table I. All the values of the correlation measure for
these typical states are in the form of $\ln d$ with $d$ a positive
integer ($0=\ln 1$), and the values in the second states are $\ln2$
larger than the values in the third states. Let us explain these
results as follows.

Because the reduced $1$-particle reduced density matrices for the
second states and the third states are the same, the maximum entropy
states $\sigma_{1}^{(2)}$ must be the same for these two states.
Thus the difference of the correlation measures for these two states
is equal to the difference of the entropies of these two states,
which is equal to $\ln2$.

The first two states in Table I are pure states, so the correlation
measures are equal to the entropy of the corresponding maximum
entropy states $\sigma^{(2)}_{1}$. As is well-known, the maximum
entropy of a state in an $m$ dimensional Hilbert space is equal to
$\ln m$. In fact, we find that the dimension of Hilbert space for
the maximum entropy state constrained by the $1$-particle reduced
density matrix is $d$ exactly. It is the existence of correlation
that makes the sate only occupy $1$-dimensional Hilbert space. So
the degree of correlation, \textit{i.e.}, the uncertainty decrease
induced by the correlation,  is $\ln d$.

\textit{Discussions and summary.} --- The correlation measure
defined by Eqs. (\ref{eq1}-\ref{eq4}) has four obvious advantages.
First, it is a universal correlation measure for all kinds of
two-particle states, that is, it is valid for bosons or fermions,
for indistinguishable particles or  identical particles, and for
pure or mixed states. Second, it gives a simple physical picture for
the correlation. The degree of correlation is measured by the amount
of uncertainty decrease induced by the correlation. In other words,
a state with more correlation has less uncertainty, and  the
uncorrelated state for a two-particle state is the state with the
same $1$-particle state and the maximum entropy simultaneously.
Third, it is computable for arbitrary two-particle states. Four, the
approach can be directly generalized to treat with the correlations
in more than two identical particles, which has been made for
distinguishable particles in Refs. \cite{Linden,Zhou2}.


In summary, we propose a correlation measure for states  of two
identical particles by using the maximum entropy principle. We
obtain the analytical results of the correlation measure, which make
it computable for arbitrary two-particle states. Based on the
correlation measure, we show that the degrees of correlation in the
same two-particle states with different particle types will decrease
in the following order: bosons, fermions, distinguishable particles.
We hope that this informative picture for correlation is helpful for
the characterization of intrinsic correlations in the system of
identical particles, and can improve our understandings on its
strongly correlated physics.

\vskip 0.5cm

The author thanks B. Zeng, L. You and C.P. Sun for helpful
discussions. This work is supported by NSF of China under Grant No.
10775176, and NKBRSF of China under Grant Nos. 2006CB921206 and
2006AA06Z104.

\end{document}